\documentclass[runningheads]{llncs}
\usepackage[T1]{fontenc}
\usepackage{graphicx}

\usepackage{xcolor}
\usepackage{multirow}
\usepackage{graphicx}
\usepackage{hyperref}
\usepackage{url}
\usepackage[ruled]{algorithm2e}
\usepackage{algorithmic} 
\usepackage{booktabs}

\usepackage{wrapfig}
\usepackage{amsmath}
\begin{document}
\title{Unlearnable Examples Detection via Iterative Filtering}
\author{Anonymous submission}
\institute{}
%
%
\author{Yi Yu\inst{1}\thanks{equal contribution} \and
Qichen Zheng\inst{1\star} \and
Siyuan Yang\inst{1}\thanks{Corresponding author}\and
Wenhan Yang\inst{2}\and
Jun Liu\inst{3}\and
Shijian Lu\inst{1}\and
Yap-Peng Tan\inst{1}\and
Kwok-Yan Lam\inst{1}\and
Alex Kot\inst{1}}
\authorrunning{Y. Yu et al.}
%
\institute{Nanyang Technological University, Singapore \and
PengCheng Laboratory, Shenzhen, China\and
Lancaster University, UK\\
\email{\{yuyi0010, qichen001, siyuan.yang, Shijian.Lu, eyptan, kwokyan.lam, eackot\}@ntu.edu.sg, yangwh@pcl.ac.cn, j.liu81@lancaster.ac.uk}} 

\maketitle              
\begin{abstract}
Deep neural networks are proven to be vulnerable to data poisoning attacks. Recently, a specific type of data poisoning attack known as availability attacks, has led to the failure of data utilization for model learning by adding imperceptible perturbations to images. Consequently, it is quite beneficial and challenging to detect poisoned samples, also known as Unlearnable Examples (UEs), from a mixed dataset. In response, we propose an Iterative Filtering approach for UEs identification. This method leverages the distinction between the inherent semantic mapping rules and shortcuts, without the need for any additional information. We verify that when training a classifier on a mixed dataset containing both UEs and clean data, the model tends to quickly adapt to the UEs compared to the clean data. Due to the accuracy gaps between training with clean/poisoned samples, we employ a model to misclassify clean samples while correctly identifying the poisoned ones. The incorporation of additional classes and iterative refinement enhances the model's ability to differentiate between clean and poisoned samples. Extensive experiments demonstrate the superiority of our method over state-of-the-art detection approaches across various attacks, datasets, and poison ratios, significantly reducing the Half Total Error Rate (HTER) compared to existing methods.

\end{abstract}
\section{Introduction}
{
The recent emphasis on data-centric AI \cite{zha2023data} underscores the importance and effectiveness of improving the model's performance from the perspective of data pre-processing optimization instead of solely the model design.
It becomes critical to address the challenges tied to those data-centric issues, including data collection, data preprocessing, data quality maintenance, \textit{etc.}, for learning-based methods.
In fact, a considerable portion of the machine learning process has been dedicated to the data issue~\cite{whang2023data}.
These efforts have gradually guided researchers towards a consensus: high-quality data is critical for enabling more advanced machine-learning algorithms to reach their full potential, which poses the data on par with the approach itself.
However, prevalent challenges persist, given that many real-world datasets are limited in size, dirty \cite{frenay2013classification}, biased \cite{tommasi2017deeper}, and in some cases, even contaminated with malicious intents \cite{shafahi2018poison}.
}

Together with the impressive performance of the deep neural networks, many concerns have been raised about their related AI security issues~\cite{Yu_2022_CVPR,wang2024benchmarking,xia2024mitigating,Yu_2023_CVPR}.
Data poisoning attacks \cite{huang2020metapoison,chen2017targeted,liu2020reflection} further intensify these difficulties and reveal the urgent requirement for the attention of the data management community.
{The significance of this issue is rooted in the deliberate actions of attackers who maliciously manipulate data to undermine the accuracy of AI applications.
Compared to the natural degradation of signals or shifts in features, these contaminations are more insidious and can result in greater damage, leading to a more obvious performance drop.}
{In this work, we focus on a specific type of data poisoning attack termed unlearnable examples (UEs) \cite{LSP,OPS,lin2024safeguarding,yu2024purify,Meng2024SemanticDH,wang2024eclipse}.
These attacks add nearly invisible perturbations to the training data, which makes the trained models fail to obtain useful knowledge for reaching reasonable performance.
}

{
The widely employed dataset search engines \cite{brickley2019google} have heightened the risks of potential threats and misuse.
Malicious dataset providers might release metadata to the public, which can be automatically discovered and propagated through search engines.
UEs, by introducing invisible perturbations to the original image, present a considerable challenge in distinguishing between clean and poisoned samples.
The malicious dataset owner could even create mixed data by poisoning only portions of the data, which further makes the detection more challenging. 
Researches~\cite{EM,AR} demonstrate that incorporating even a small proportion of clean samples into an unlearnable dataset leads to an increase in test accuracy. 
As a result, it is difficult for data users to determine whether the dataset is normal because the trained model appears to have reasonable performance.
Training on these UEs may not fully leverage the model's potency, compromising testing performance and wasting computational resources and time.
}

{
In this paper, we introduce Iterative Filtering (IF), an effective detection technique tailored for detecting a wide range of visually imperceptible UEs.
Our analysis reveals a critical insight: when a classifier is trained on a dataset that combines UEs with clean data, the model tends to adapt more rapidly to the UEs than to the clean data.
This suggests that when the model is evaluated on previously unseen UE, it often demonstrates superior accuracy compared to its performance on clean samples.
Exploiting the accuracy disparities between training on clean and poisoned samples, we design a model to misclassify clean samples while accurately identifying the poisoned ones, facilitating the detection of tainted data.
Furthermore, our findings lead us to the formulation of the Iterative Filtering algorithm.
With additional classes and iterative refinement, the proposed approach achieves improved performance in distinguishing these two types of samples.
}
Our contributions can be summarized as:
\begin{itemize}
\vspace{-2mm}
    \item In this paper, we address a critical issue in the era of deep learning: how to identify and filter harmful unlearnable data, screening samples that cannot be learned from. Accordingly, we introduce Iterative Filtering (IF), the first detection method that aims at identifying visually imperceptible UEs.
    \item  
    {
    IF capitalizes on the observation that models trained on datasets blending UEs with clean data tend to adjust more swiftly to UEs, resulting in a discernible accuracy differential, to enable the identification of UEs.
    Specifically, IF integrates additional classes and undergoes iterative refinement, enhancing its discrimination between clean and poisoned samples.
    }
    \item Extensive experiments show the superior performance of our method over SOTA detection approaches when confronted with various types of attacks, datasets, and poison ratios. 
    When employing a detection-purification strategy, the results further emphasize the method's robustness in strengthening defenses against UEs.
\end{itemize}

\section{Related Work}
\subsection{Data poisoning}
Data poisoning~\cite{biggio2012poisoning,hong2020effectiveness,huang2020metapoison,koh2022stronger} is an increasingly recognized challenge in the modern machine learning ecosystem. Essentially, it involves the malicious modification of training data (often in a passive manner) to deliberately distort the behavior of a machine learning model. 
{
These data poisoning attacks manifest in multiple forms, ranging from targeted attacks aimed at particular categories to UEs seeking to undermine overall test accuracy.
For example, Backdoor attacks~\cite{chen2017targeted,doan2021lira,gu2019badnets,nguyen2021wanet} are characterized by the manipulation of training data instances. This allows attackers to control the target model's output utilizing a predetermined trigger.
}

While most of the literature focuses on the malicious use of poisoning attacks, UEs are employed as a protective measure against unauthorized model training. 
For example, Error-Minimizing (EM)~\cite{EM} poisons introduce error minimization noise to prevent deep learning models from absorbing knowledge. 
Targeted Adversarial Poisoning (TAP)~\cite{TAP} uses targeted adversarial examples of pre-trained models for UEs. 
Robust Error-Minimizing (REM)~\cite{REM} introduces adversarial training based on the EM method to generate robust UEs. 
Self-ensemble protection (SEP)~\cite{chen2022self} uses multiple model checkpoints' gradients to generate poison in a self-ensemble manner.
Linear separable Synthetic Perturbation(LSP)~\cite{LSP} reveals that the perturbations of several existing UEs are (almost) linearly separable and propose to use synthetic shortcuts to perform availability attack.
Recently, 
{One Pixel Shortcut (OPS)~\cite{OPS} delves into the model's susceptibility to sparse poisons and augments its resistance to adversarial training.}

\subsection{Existing Defense against UEs}
Defense against UEs can be mainly classified into two distinct categories: preprocessing and training-phase defenses. 
The method based on preprocessing eliminates the poison added by the attacker by preprocessing the data before training. Recently, \cite{liu2023image} proposes to purify the poisoned data using grayscale transformation and JPEG compression~\cite{marcellin2000overview}. \cite{AVATAR} demonstrates the efficacy of diffusion models~\cite{ho2020denoising} in removing data protection perturbations. 
{Training-phase defense methods are characterized by alterations in the training procedure to enable model robustness, even when exposed to poisoned data.}
Existing work tends to adopt adversarial training~\cite{pgd} as the countermeasure.
{However, this approach is not without its shortcomings – the substantial computational overhead and extended training durations often outweigh its benefits, and there exists a tangible risk of compromising the performance of models trained on clean datasets. }
{Recently, adversarial augmentations~\cite{qin2023learning} introduces an innovative technique of applying a spectrum of augmentations.

\subsection{Detection of Backdoor attacks}
A similar but different task compared to availability attack detection is backdoor attack detection. 
To date, some defense methods have been proposed to detect and mitigate backdoor attacks.
{
\cite{steinhardt2017certified} unveils a general defense against poisoning attacks, leveraging outlier or anomaly detection techniques. 
However, a significant limitation of their approach is the prerequisite of a clean, trusted dataset to effectively train the outlier detector. 
Addressing this constraint, \cite{chen2018detecting} introduces a pioneering methodology capable of detecting poisonous backdoors without the necessity of a verified and trusted dataset.
This approach involves analyzing the neural network activations associated with the training data. It assesses whether the data is poisoned and identifies the specific data points that are affected if poisoning is detected.
} 
\cite{tran2018spectral} proposes a Spectral Signature defense that removes the data with the top $\epsilon$ eigenvalue.
\cite{wang2019neural} proposes a Neural Cleanse defense that first reverse-engineers a trigger by searching for patches that cause strong misclassification, then prunes neurons with large activations. 
\cite{peri2020deep} capitalizes on the observation that adversarial examples exhibit distinct feature distributions in higher layers of a neural network compared to their clean counterparts. 
They introduce a straightforward yet remarkably effective defense mechanism called Deep k-NN, which is designed for detecting and removing poisoned samples by leveraging these differences in feature distributions.
Recently, \cite{liu2023detecting} introduces TeCo, a technique anchored in sample corruption consistency for the precise detection of trigger samples during testing.

\section{Preliminary}

\textbf{Unlearnable Examples.} 
{
In the era of big data, the internet and search engines bring about massive volumes of data that can be used for training deep models.
However, there is a possibility of data contamination caused by unlearnable examples.
That is, with small imperceptible perturbations, the data appears normal still in visual appearance but results in the failure of training deep networks with unsatisfactory testing accuracy.
Here, these perturbed samples are termed `unlearnable examples' (UEs).
}
Given a clean dataset $\mathcal{D}_c=\{\left({x}_i, y_i\right)\}_{i=1}^N$ comprising $N$ training samples, 
where ${x} \in \mathcal{X} \subset {R}^d$ represents the image and $y \in \mathcal{Y}=\{0, \cdots, C-1\}$ {represents its labels.}
We consider a scenario {in which} a classifier is trained on the unlearnable data, represented as $f_{\theta}: \mathcal{X} \rightarrow \mathcal{Y}$.
To corrupt the model training,
{the existing methods introduce perturbations to the clean images, resulting in an unlearnable dataset defined as:}
\vspace{-1mm}
\begin{equation}
    \mathcal{D}_u=\left\{\left({x}_i+{\delta}_i, y_i\right)\right\}_{i=1}^N,
\end{equation}
where ${\delta}_i \in \Delta_{\mathcal{D}} \subset {R}^d$, with $\Delta_{\mathcal{D}}$ representing the perturbation set for $\mathcal{D}_c$.
The objective of unlearnability is to ensure that a classifier $f_\theta$ trained on $\mathcal{D}_u$ exhibits poor performance on testing datasets when using $\mathcal{D}_u$ in the inference stage.

\noindent \textbf{UEs Detection.}
Given a mixed dataset $\mathcal{D} = \mathcal{D}_u \cup \mathcal{D}_c$, where  $\mathcal{D}_u = \{({x}_i+{\delta}_i,y_i)\}_{i=1}^{N_u}$ consists of ${N_u}$ UEs and $\mathcal{D}_c = \{(x_i,y_i)\}_{i=1}^{N_c}$ including ${N_c}$ clean samples, 
the detection of UEs within the mixed dataset $\mathcal{D}$ turns to learn a mapping $f$ appropriate for a binary classification problem, relative to the poison ratio $p=\frac{N_u}{N_u+N_c}$.
Mathematically, the problem can be formulated as: 
\vspace{-2mm}
\begin{equation}
\min_{f(x) \in \{0,1\}} \sum_{i=1}^{N_u+N_c}|f(x)-I\{x \in \mathcal{D}_u\}|.
\end{equation}

\section{Methodology}
\subsection{Key Intuition}
\noindent \textbf{UEs make faster learners.} 
{Recent research indicates that UEs offer easily learnable features that are closely linked to labels, commonly recognized as shortcuts~\cite{LSP,AR}.}
{A notable observation is that when training a classifier on a dataset merging UEs and clean data, the model adapts to the UEs more quickly than it does to the clean data~\cite{EM,TAP,LSP}.}
Basically, when training a classifier $F(\cdot|\theta)$ on $\mathcal{D} = \mathcal{D}_u \cup \mathcal{D}_c$, the optimization process (regarding the loss $\mathcal{L}$) is given by
\vspace{-2mm}
\begin{equation}
        \theta = \arg\min_{\theta} \sum_{i=1}^{N_u}\mathcal{L}(F(x_i+\delta_i|\theta),y_i)+\sum_{i=1}^{N_c}\mathcal{L}(F(x_i|\theta),y_i).
\end{equation}

\vspace{-2mm}
Since $\delta_i$ serves as a shortcut, the optimization process will prioritize minimizing the loss on UEs over clean data. This indicates that when evaluated on previously {unseen UE}, which typically follow a similar distribution to the UEs in the training set~\cite{liu2023image,LSP,OPS},
{the model often shows superior accuracy compared to its performance on clean data.}

To validate this phenomenon, we consider a mixed dataset $\mathcal{D}$ that includes an equal number of UEs and clean data. 
Subsequently, we randomly sample 50\% of the data from $\mathcal{D}$ for classifier training and evaluate on the remaining data. 
{Figure~\ref{figure1} illustrates the testing accuracy for both the unseen UEs and the clean data throughout each epoch.} 
{Notably, for EM~\cite{EM} and OPS~\cite{OPS}, the accuracy for UEs increases near 100\% within a handful of epochs, while accuracy improvement for the clean data is relatively slow.}

{This observation suggests that leveraging the distinction in learnability between UEs and clean data, especially through the introduction of an \textbf{early stopping} during training, can help to filter out potential inaccuracies labeled as clean data.}
In each iteration, we randomly select training data from the mixed dataset and evaluate on the remaining data. 
Through this iterative process, we can progressively distinguish the clean data from the rest.
Nevertheless, when clean data is part of the optimization process as depicted in Eq.~\ref{optimization}, the accuracy for such data might not necessarily remain at a low value.
Consequently, if there is no significant gap in accuracy between UEs and clean data, the filtering-based detection approach may fail to obtain the desirable results.

\begin{figure}[t]
\begin{minipage}{1.0\linewidth}
\centerline{{\includegraphics[width=1.0\linewidth]{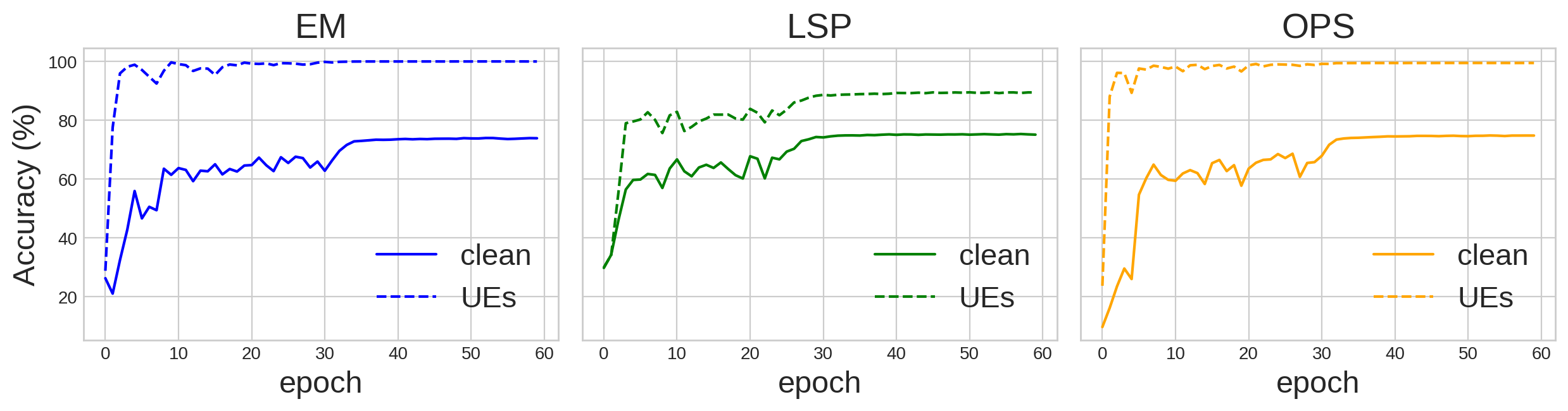}}}
\end{minipage}
\vspace{-6mm}
    \caption{Test accuracy (\%) on the unseen unlearnable data and clean data when training a classifier on a mixed dataset.}
    \label{figure1}
    \vspace{-3mm}
\end{figure}

\begin{figure}[t]
\begin{minipage}{1.0\linewidth}
\centerline{{\includegraphics[width=1.0\linewidth]{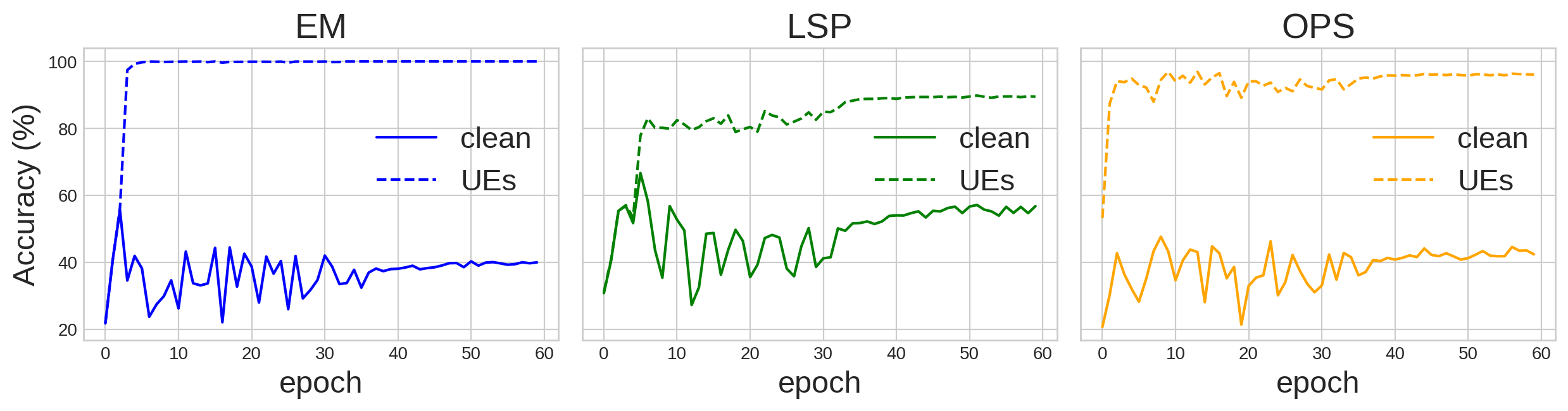}}}
\end{minipage}
\vspace{-6mm}
    \caption{Test accuracy (\%) on the unseen unlearnable data and clean data when training a classifier on a mixed dataset plus additional clean data with label set to $y+C$.}
    \label{figure2}
\vspace{-3mm}
\end{figure}

\begin{figure}[t]
\begin{minipage}{1.0\linewidth}
{{\includegraphics[width=0.96\linewidth]{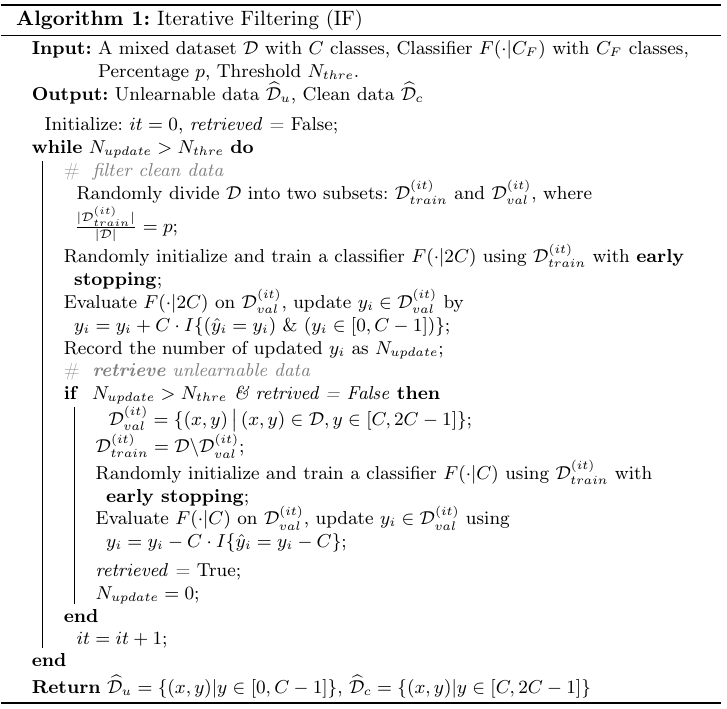}}}
\end{minipage}
\label{filtering}
\vspace{-6mm}
\end{figure}

\begin{figure}[t]
\begin{minipage}{1.0\linewidth}
\centerline{{\includegraphics[width=1.0\linewidth]{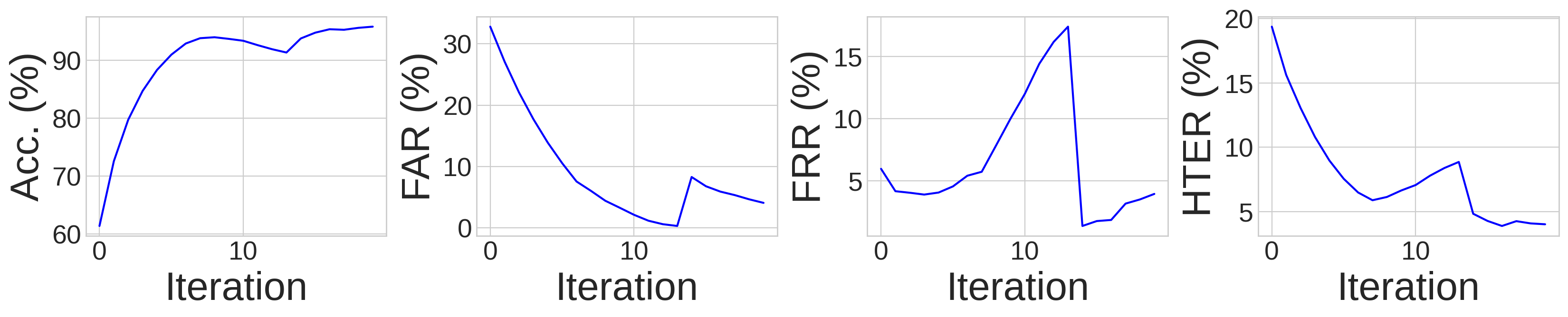}}}
\end{minipage}
\vspace{-5mm}
    \caption{Performance Vs. Iterations on detecting EM with 80\% poison ratio.}
    \label{figure3}
\vspace{-3mm}
\end{figure}

\textbf{It is simpler to distinguish between UEs and clean data.} 
To tackle the above-mentioned issue, we reveal a key insight: 
{Distinguishing between clean and poisoned samples is demonstrated to be less challenging than guiding the model to misclassify clean samples while accurately classifying poisoned ones.}
In the filtering-based algorithm described above,
{the clean data, once filtered, is excluded from model training in subsequent iterations.
Moreover, our studies provide evidence suggesting that these excluded clean data can considerably improve the effectiveness of detection.
}

{Considering a mixed training dataset comprising 50\% UEs and 50\% clean data, each with their true labels $y$. 
Additionally, we have the excluded clean data, but in this case, we modify their labels to $y+C \in \{C, \cdots, 2C-1\}$, where $C$ represents the number of classes in the dataset. 
These newly labeled data are then merged with the mixed dataset to form the dataset for the next iteration.
The improved version of the optimization process is as follows:
\vspace{-3mm}
\begin{equation}
        \theta = \arg\min_{\theta} \sum_{i=1}^{N_u}\mathcal{L}(F(x_i+\delta_i|\theta),y_i)+\sum_{i=1}^{N_c}\mathcal{L}(F(x_i|\theta),y_i)\\
        +\sum_{i=1}^{N_c^{'}}\mathcal{L}(F(x_i|\theta),y_i+C).
    \label{optimization}
\end{equation}

\vspace{-2mm}
Subsequently, we randomly select 50\% of the data for training, utilize the expanded label space $\{0, \cdots, 2C-1\}$, and evaluate the remaining data with original labels $y \in\{0, \cdots, C-1\}$.}
{
As illustrated in Figure~\ref{figure2}, there is a noticeable decrease in the accuracy of clean data compared to previous experiments. 
This result is intuitive, as introducing clean data with additional class labels can lead to these clean data points being misclassified as $y+C$, thereby amplifying the difference between UEs and clean data.
Moreover, as the number of iterations increases, the percentage of clean data with modified labels is expected to rise, likely resulting in more misclassifications as $y+C$, thereby further expanding the gap between UEs and clean data.
}

\subsection{Iterative Filtering (IF) for UEs Detection}
{
The Iterative Filtering, as outlined in Algorithm~\ref{filtering}, is designed to distinguish between UEs and clean samples, employing one subset for training and the other for testing and classification.
Specifically, if a sample is incorrectly identified during testing, its class label is adjusted to $y+C$, expanding the label space for the entire dataset to $\{0, \cdots, 2C-1\}$.
Consequently, we introduce $C$ additional classes to the original classifier.
}

In essence, our approach leverages the inherent responses of models when they encounter UEs.
{Through iterative training and testing, and by categorizing samples based on the model's prediction capabilities, we can effectively distinguish the UEs and clean samples within the dataset.}
During the iterative filtering process, poison samples tend to cluster within the initial $C$ categories, whereas clean samples are primarily categorized within the subsequent $C$ classes.
{
However, in the early iterations, given a relatively low proportion of UEs in the training data, there exists a possibility that some UEs might be incorrectly identified as clean data during testing. 
As illustrated in Figure~\ref{figure3}, as iterations increase, there is a corresponding rise in the FRR, indicating a growing number of poisoned samples being misclassified with each subsequent iteration.
To address this, we propose the idea of conducting retrieval once after several iterations. 
Furthermore, we establish a stopping criterion based on the number of correctly classified clean data during each iteration.
Conventionally, we set the threshold $N_{\text{thre}}$ to be 2\% of the total number of the entire dataset, and $p$ to be 50\% for splitting in each iteration.
}

\section{Experiments}
\subsection{Experimental Setup}
\textbf{Datasets and models.} 
We use three image classification datasets, CIFAR-10~\cite{cifar10}, CIFAR-100~\cite{cifar10}, and 100-class subset of ImageNet~\cite{imagenet} in our experiments. 
We implement ResNet-18~\cite{resnet18} as the image classification model.

\noindent \textbf{Methods for Generating UEs.} 
{We explore various representative methods to generate UEs: EM~\cite{EM}, REM~\cite{REM} with a perturbation bound of $\ell_{\infty}=\frac{8}{255}$, AR~\cite{AR}, LSP~\cite{LSP} constrained by $\ell_{2}=1.0$, and OPS~\cite{OPS} constrained by $\ell_{0}=1$.
We adhere to the original papers to determine these bounds across various norms, ensuring that all introduced perturbations maintain uniform noise levels.
}

\noindent \textbf{Poison Ratio (R).} We select the ratio from $\{20,40,60,80\}\%$ to demonstrate the effectiveness of our methods across different poisoning ratios. 

\noindent \textbf{Competing Methods.} 
{As our proposed IF is the first UEs detection method without any baselines, 
we re-implement several state-of-the-art defensive techniques from backdoor attacks for comparison.}
TeCo~\cite{liu2023detecting} is an innovative test-time poisoned sample detection method designed for backdoor attacks. 
It evaluates the consistency of test-time robustness by measuring the extent of severity deviation, which triggers shifts in predictions across different corruptions.
Deep k-NN~\cite{peri2020deep} is a straightforward, yet highly-effective detection against clean-label backdoor attacks, and exploits the property that adversarial examples have different feature distributions than their clean counterparts in {deeper} 
layers of the network.
{Furthermore, these adversarial features are typically aligned closely to the distribution of the target class.}

\noindent \textbf{Metrics.}
{To evaluate the performance of UEs detection, we adopt the Half Total Error Rate (HTER) and the accuracy (Acc.) as metrics. 
HTER, the False Rejection Rate (FRR) and the False Acceptance Rate (FAR) are given by:
\vspace{-2mm}
\begin{equation}
\text{HTER}=\frac{\text{FRR}+\text{FAR}}{2},
    ~\text{FAR}=\frac{\text{FP}}{\text{FP}+\text{TN}}, ~\text{FRR}=\frac{\text{FN}}{\text{FN}+\text{TP}}.
\end{equation}
%
HTER integrates both FRR and FAR to provide a holistic view.  

\begin{table}[t]
\caption{Detection performance (\%) on CIFAR-10 dataset.}
\vspace{-3mm}
\centering
\scalebox{0.85}{
\setlength{\tabcolsep}{0.8pt}
\begin{tabular}{c|c|*4{c}|*4{c}|*4{c}}
\toprule
\multirow{2}{*}{\shortstack{R\\(\%)}} & \multirow{2}{*}{UEs} &
\multicolumn{4}{c|}{Deep k-NN~\cite{peri2020deep}} &
\multicolumn{4}{c|}{TeCo~\cite{liu2023detecting}} &
\multicolumn{4}{c}{Our Method} \\
\cmidrule{3-14}
&& Acc. $\uparrow$ & FAR $\downarrow$ & FRR $\downarrow$ & HTER $\downarrow$ & Acc. $\uparrow$ & FAR $\downarrow$ & FRR $\downarrow$ & HTER $\downarrow$ & Acc. $\uparrow$ & FAR $\downarrow$ & FRR $\downarrow$ & HTER $\downarrow$ \\
\midrule
\multirow{5}{*}{20} & EM & 47.7 & 3.8 & 73.4 & 38.6 & 49.6 & 19.5 & 79.6 & 49.5 & 83.3 & 5.7 & 17.1 & 11.4 \\
& OPS & 46.3 & 2.3 & 73.4 & 37.8 & 51.5 & 19.9 & 79.9 & 49.9 & 91.1 & 4.7 & 9.5 & 7.1 \\
& LSP & 38.0 & 16.5 & 78.3 & 47.4 & 49.5 & 19.6 & 79.7 & 49.6 & 79.3 & 60.7 & 19.3 & 40.0 \\
& REM & 38.2 & 16.3 & 78.4 & 47.4 & 48.5 & 20.2 & 80.2 & 50.2 & 79.4 & 72.6 & 19.9 & 46.2 \\
& AR & 38.6 & 19.3 & 79.6 & 49.5 & 51.4 & 19.7 & 79.6 & 49.7 & 78.7 & 78.0 & 20.0 & 49.0 \\
\cmidrule{1-14}
\multirow{5}{*}{40} & EM & 69.9 & 3.3 & 42.9 & 23.1 & 49.5 & 40.3 & 60.2 & 50.3 & 88.9 & 0.1 & 15.6 & 7.9 \\
& OPS & 61.0 & 21.5 & 49.4 & 35.4 & 48.5 & 40.5 & 60.4 & 50.5 & 92.9 & 0.6 & 10.0 & 5.4 \\
& LSP & 64.2 & 12.1 & 46.7 & 29.9 & 49.7 & 39.5 & 59.6 & 49.6 & 80.2 & 3.0 & 24.4 & 13.7 \\
& REM & 57.2 & 27.0 & 52.3 & 39.7 & 49.5 & 39.9 & 59.9 & 49.9 & 82.2 & 4.6 & 22.1 & 13.3 \\
& AR & 48.1 & 36.8 & 58.3 & 47.6 & 48.8 & 40.1 & 60.1 & 50.1 & 60.0 & 43.0 & 40.0 & 41.5 \\
\cmidrule{1-14}
\multirow{5}{*}{60} & EM & 74.4 & 29.5 & 23.3 & 26.4 & 51.1 & 59.6 & 39.7 & 49.7 & 95.9 & 4.1 & 3.9 & 4.0 \\
& OPS & 70.8 & 32.4 & 6.2 & 19.3 & 50.3 & 60.0 & 40.0 & 50.0 & 95.3 & 2.4 & 8.0 & 5.2 \\
& LSP & 60.0 & 50.4 & 30.6 & 40.5 & 49.7 & 61.4 & 41.1 & 51.2 & 94.0 & 4.3 & 8.5 & 6.4 \\
& REM & 68.8 & 35.9 & 29.1 & 32.5 & 49.5 & 60.6 & 40.6 & 50.6 & 85.0 & 16.0 & 13.0 & 14.5 \\
& AR & 40.5 & 59.5 & 39.8 & 49.6 & 50.3 & 60.1 & 40.1 & 50.1 & 40.8 & 14.1 & 59.7 & 36.9 \\
\cmidrule{1-14}
\multirow{5}{*}{80} & EM & 43.2 & 78.7 & 17.6 & 48.1 & 51.2 & 79.8 & 19.8 & 49.8 & 98.1 & 2.1 & 0.9 & 1.5 \\
& OPS & 70.5 & 35.6 & 25.9 & 30.7 & 50.1 & 80.4 & 20.4 & 50.4 & 96.0 & 0.1 & 16.5 & 8.3 \\
& LSP & 75.6 & 57.0 & 10.1 & 33.5 & 52.7 & 81.0 & 20.8 & 50.9 & 96.1 & 3.2 & 7.0 & 5.1 \\
& REM & 58.8 & 72.5 & 13.2 & 42.9 & 56.6 & 79.9 & 20.0 & 50.0 & 90.5 & 8.8 & 13.4 & 11.1 \\
& AR & 49.7 & 79.8 & 19.9 & 49.9 & 49.4 & 80.7 & 20.7 & 50.7 & 82.4 & 17.2 & 26.2 & 21.7 \\
\cmidrule{1-14}
\multicolumn{2}{c|}{Average} & 56.1 & 34.5 & 42.4 & 38.5 & 50.4 & 50.1 & 50.1 & 50.1 & \textbf{84.5} & \textbf{17.3} & \textbf{17.8} & \textbf{17.5} \\
\bottomrule
\end{tabular}
}
\label{table:exp_cifar10}
\vspace{-3mm}
\end{table}

\begin{table}[t]
\caption{Performance (\%) on CIFAR-100 dataset.}
\vspace{-3mm}
\centering
\scalebox{0.85}{
\setlength{\tabcolsep}{0.8pt}
\begin{tabular}{c|c|*4{c}|*4{c}|*4{c}}
\toprule
\multirow{2}{*}{\shortstack{R\\(\%)}} & \multirow{2}{*}{UEs} &
\multicolumn{4}{c|}{Deep k-NN~\cite{peri2020deep}} &
\multicolumn{4}{c|}{TeCo~\cite{liu2023detecting}} &
\multicolumn{4}{c}{Our Method} \\
\cmidrule{3-14}
&& Acc. $\uparrow$ & FAR $\downarrow$ & FRR $\downarrow$ & HTER $\downarrow$ & Acc. $\uparrow$ & FAR $\downarrow$ & FRR $\downarrow$ & HTER $\downarrow$ & Acc. $\uparrow$ & FAR $\downarrow$ & FRR $\downarrow$ & HTER $\downarrow$ \\
\midrule
\multirow{3}{*}{20}&EM&52.1&10.8&73.5&42.2&45.8&19.9&80.0&50.0&81.3&37.9&17.6&27.7\\
&OPS&50.8&10.2&73.6&41.9&48.1&19.9&79.9&49.9&80.9&37.8&18.4&28.1 \\
&LSP&45.5&17.8&78.4&48.1&47.7&19.9&79.9&49.9&79.2&79.7&20.0&49.8\\
\cmidrule{1-14} 
\multirow{3}{*}{40}&EM&73.1&2.0&40.0&21.0&48.3&40.5&60.4&50.5&93.0&4.2&8.5&6.4\\
&OPS&69.7&5.5&42.7&24.1&48.9&40.9&60.8&50.8&90.4&2.9&12.7&7.8 \\
&LSP&66.9&8.3&44.7&26.5&45.9&42.6&61.8&52.2&78.6&11.7&24.5&18.1\\
\cmidrule{1-14}
\multirow{3}{*}{60}&EM&90.3&2.9&12.9&7.9&50.8&59.9&39.9&49.9&96.9&4.0&1.7&2.9\\
&OPS&68.8&3.7&43.6&23.7&49.7&61.0&40.9&50.9&92.1&1.7&15.2&8.5\\
&LSP&85.3&7.4&17.8&12.6&50.6&60.3&40.2&50.3&92.8&6.3&8.6&7.4\\
\cmidrule{1-14}
\multirow{3}{*}{80}&EM&42.3&75.7&9.0&42.3&57.0&81.0&20.6&50.8&97.5&2.6&1.8&2.2\\
&OPS&71.2&60.3&5.2&32.8&49.5&81.0&20.9&50.9&96.1&2.0&11.4&6.7\\
&LSP&71.8&59.4&4.8&32.1&56.8&77.4&18.1&47.7&97.1&1.6&8.1&4.9 \\
\cmidrule{1-14}
\multicolumn{2}{c|}{Average} & 59.2 & 33.2 & 41.5 & 37.4 & 52.2 & 47.5 & 50.2 & 49.3 & \textbf{93.7} & \textbf{11.5} & \textbf{4.1} & \textbf{7.9} \\
\bottomrule
\end{tabular}
}
\label{table:exp_cifar100}
\vspace{-3mm}
\end{table}

\begin{table}[t]
\caption{Performance (\%) on 100-class ImageNet subset.}
\vspace{-3mm}
\centering
\scalebox{0.85}{
\setlength{\tabcolsep}{0.8pt}
\begin{tabular}{c|c|*4{c}|*4{c}|*4{c}}
\toprule
\multirow{2}{*}{\shortstack{R\\(\%)}} & \multirow{2}{*}{UEs} &
\multicolumn{4}{c|}{Deep k-NN~\cite{peri2020deep}} &
\multicolumn{4}{c|}{TeCo~\cite{liu2023detecting}} &
\multicolumn{4}{c}{Our Method} \\
\cmidrule{3-14}
&& Acc. $\uparrow$ & FAR $\downarrow$ & FRR $\downarrow$ & HTER $\downarrow$ & Acc. $\uparrow$ & FAR $\downarrow$ & FRR $\downarrow$ & HTER $\downarrow$ & Acc. $\uparrow$ & FAR $\downarrow$ & FRR $\downarrow$ & HTER $\downarrow$ \\
\midrule
\multirow{3}{*}{20} & EM & 41.4 & 19.9 & 80.1 & 50.0 & 49.8 & 20.3 & 80.2 & 50.2 & 93.0 & 23.6 & 1.5 & 12.6 \\
& OPS & 42.3 & 18.2 & 79.1 & 48.6 & 50.3 & 20.8 & 80.1 & 50.9 & 90.1 & 26.6 & 5.3 & 16.0 \\
& LSP & 49.9 & 6.6 & 73.0 & 39.8 & 50.2 & 17.9 & 78.1 & 48.0 & 88.8 & 22.8 & 9.0 & 15.9 \\
\cmidrule{1-14} 
\multirow{3}{*}{40} & EM & 47.1 & 40.3 & 59.9 & 50.1 & 48.6 & 40.9 & 60.8 & 50.8 & 94.0 & 11.4 & 1.7 & 6.5 \\
& OPS & 60.2 & 18.0 & 49.9 & 33.9 & 49.0 & 41.1 & 61.1 & 51.1 & 91.7 & 12.7 & 5.0 & 8.8 \\
& LSP & 68.6 & 4.2 & 43.6 & 23.9 & 55.1 & 34.5 & 54.8 & 44.6 & 92.0 & 15.4 & 1.6 & 8.5 \\
\cmidrule{1-14}
\multirow{3}{*}{60} & EM & 52.9 & 60.3 & 39.8 & 50.1 & 48.9 & 62.0 & 41.7 & 51.9 & 95.8 & 6.3 & 0.1 & 3.4 \\
& OPS & 77.9 & 23.2 & 21.5 & 23.4 & 48.5 & 61.2 & 41.3 & 51.3 & 94.6 & 4.8 & 6.2 & 5.5 \\
& LSP & 85.9 & 3.9 & 18.1 & 11.0 & 56.7 & 54.0 & 35.8 & 44.9 & 95.7 & 4.8 & 3.5 & 4.1 \\
\cmidrule{1-14}
\multirow{3}{*}{80} & EM & 58.8 & 79.7 & 19.7 & 49.7 & 49.3 & 81.0 & 21.0 & 51.0 & 97.2 & 3.1 & 1.0 & 2.1 \\
& OPS & 80.4 & 49.2 & 5.6 & 27.4 & 49.5 & 80.7 & 20.7 & 50.7 & 94.7 & 4.6 & 8.3 & 6.5 \\
& LSP & 44.9 & 74.9 & 7.2 & 41.0 & 57.7 & 76.0 & 17.0 & 46.5 & 96.2 & 3.2 & 6.1 & 4.7 \\
\cmidrule{1-14}
\multicolumn{2}{c|}{Average} & 59.2 & 33.2 & 41.5 & 37.4 & 52.2 & 47.5 & 50.2 & 49.3 & \textbf{93.7} & \textbf{11.5} & \textbf{4.1} & \textbf{7.9} \\
\bottomrule
\end{tabular}
}
\label{table:exp_imagenet}
\end{table}

\begin{table}[t]
\caption{Ablation study of IF on CIFAR-10. ES is the use of early stopping, and AC is the use of additional classes.}
\vspace{-3mm}
\centering
\scalebox{1.0}{
\setlength{\tabcolsep}{1.5pt}
\begin{tabular}{c|c|*2{c}|*2{c}|*2{c}|*2{c}}
\toprule
\multirow{2}{*}{UEs} & \multirow{2}{*}{Methods} &
\multicolumn{2}{c|}{40\%} &
\multicolumn{2}{c|}{60\%} &
\multicolumn{2}{c|}{80\%} &
\multicolumn{2}{c}{Average} \\
\cline{3-10}
&& Acc. $\uparrow$ & HTER $\downarrow$ & Acc. $\uparrow$ & HTER $\downarrow$ & Acc. $\uparrow$ & HTER $\downarrow$ & Acc. $\uparrow$ & HTER $\downarrow$ \\
\midrule
\multirow{4}{*}{EM} & w/o ES & \textbf{96.5} & \textbf{3.9} & 88.9 & 5.8 & 90.5 & 4.8 & 92.0 & 4.8 \\
& w/o retrieve & 94.5 & 5.2 & 91.4 & 8.9 & 97.1 & 5.1 & 94.3 & 6.4 \\
& w/o AC & 82.9 & 15.7 & 78.5 & 13.5 & 94.6 & 4.0 & 85.3 & 11.1 \\
& Ours & 88.9 & 7.9 & \textbf{96.0} & \textbf{4.0} & \textbf{98.1} & \textbf{1.5} & \textbf{94.3} & \textbf{4.5} \\
\cmidrule{1-10}
\multirow{4}{*}{OPS} & w/o ES & 94.0 & 6.4 & 92.3 & 6.7 & 93.2 & 8.9 & 93.2 & 7.3 \\
& w/o retrieve & \textbf{94.3} & \textbf{4.8} & 94.5 & 5.9 & 94.8 & 10.0 & 94.5 & 6.9 \\
& w/o AC & 83.9 & 15.9 & 87.8 & 10.7 & 92.3 & 9.6 & 88.0 & 12.1 \\
& Ours & 93.0 & 5.4 & \textbf{95.3} & \textbf{5.2} & \textbf{96.0} & \textbf{8.3} & \textbf{94.8} & \textbf{6.3} \\
\bottomrule
\end{tabular}
}
\label{table:exp_cifar10_ablation}
\vspace{-3mm}
\end{table}

\subsection{Experimental Results}
\textbf{Results on CIFAR-10 dataset.}
We evaluate the performance of IF on different UEs methods comprehensively.
The results in Table~\ref{table:exp_cifar10} demonstrate that IF can successfully identify the UEs, particularly when the poison ratio exceeds 20\%, as confirmed by the fact that most HTER valUE below 10\%.
However, {with only a 20\% poison ratio,}
the poisoned samples represent a minority, and additional experiments reveal that a marginal percentage of UEs does not significantly affect testing performance. 
{Therefore, the effectiveness of detection becomes more critical when dealing with larger poison ratios.}
{Notably, on the CIFAR-10, IF achieves an average HTER of 0.175, F1 score of 0.845, FAR of 0.173, and FRR of 0.178.
These results showcase IF's superiority, surpassing the runner-up by roughly 20\% in HTER, 30\% in Acc., 17\% in FAR, and 25\% in FRR.
}
{
Certainly, there are scenarios in which IF may not achieve success, particularly when dealing with AR.
}
AR utilize autoregressive perturbations, making them more complex and not linearly separable~\cite{AR}. 
This complexity can make them more challenging to detect, especially when the poison ratios are low.
Figure~\ref{figure3} illustrates the performance at each iteration.
It is noticeable that once retrieval is integrated and the filtering process continUE through several iterations, the method demonstrates improved convergence performance compared to the results preceding retrieval.
In summary, our work delivers consistent effectiveness across various types of UEs without using extra knowledge.

There are also some interesting findings about baselines. 
TeCo, though highly effective in detecting poisoned samples for backdoor attacks, appears to completely fail when dealing with UEs. 
This divergence in performance could be attributed to the differing objectives of backdoor attacks and UEs.
Backdoor attacks aim to manipulate the predictions, shifting them from the source class to the target class after the integration of triggers. 
On the other hand, UEs do not necessarily require a prediction shift after the perturbations are introduced.
This difference in objectives illuminates the varying performance of TeCo in these diverse situations.
Deep k-NN appears to be effective when the number of UEs and clean data is well balanced.
However, it seems to fail when dealing with a critical scenario where UEs significantly outnumber the clean data. 
This indicates that Deep k-NN's efficiency is closely tied to data distribution, 
particularly when there is a significant imbalance between UEs and clean data.

\noindent \textbf{Results on CIFAR-100 dataset and 100-class ImageNet subset.} 
In our experiments on both the 100-class datasets, we select the three most representative unlearnable examples methodologies. 
As the experimental results are shown in Table~\ref{table:exp_cifar100} and Table~\ref{table:exp_imagenet}, our IF strategy consistently outperforms competing methods in terms of detection and defense against UEs.

\subsection{Ablation Study}
{We conduct experiments to show the effectiveness of the proposed iterative filtering, comparing it with the method without early stopping, without the retrieval process and without introducing additional classes.
The experimental results are shown in Table~~\ref{table:exp_cifar10_ablation}.
The results indicate that all introduced modules in IF contribute to the enhancement of detection performance, as measured by the average metrics.
Among these strategies, the introduction of additional classes has proven to be the most effective. While early stopping may slightly impact performance at a low poison ratio, it is highly efficient, requiring lower computations, and performs well at high poison ratios.
In particular, the retrieval process has proven effective, notably in preserving high accuracy while slightly improving the HTER.
}

{
Furthermore, through t-SNE~\cite{tsne}, we illustrate the feature distribution of the model, comparing training scenarios without and with the additional clean data (having updated labels $y\in[C,2C-1]$), as shown in Figure~\ref{tsne}. 
It is evident that the additional clean data serves to separate the clean data from the UEs in the latent space, thereby improving the detection performance.
}

\begin{figure}[t]
\begin{minipage}{0.49\linewidth}
\centerline{{\includegraphics[width=1.0\linewidth]{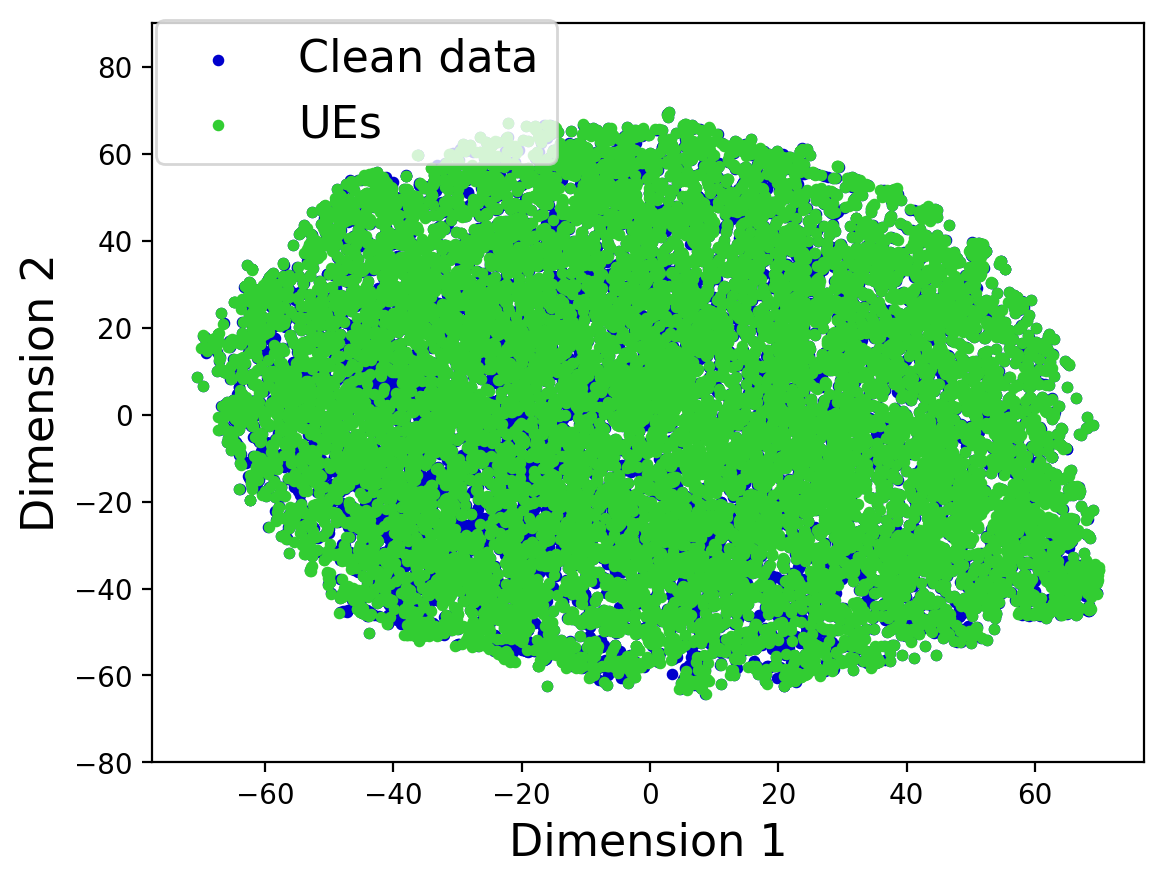}}}
\vspace{-1mm}
\centerline{(a) w/o additional clean data}
\vspace{2mm}
\end{minipage}
\begin{minipage}{0.49\linewidth}
\centerline{{\includegraphics[width=1.0\linewidth]{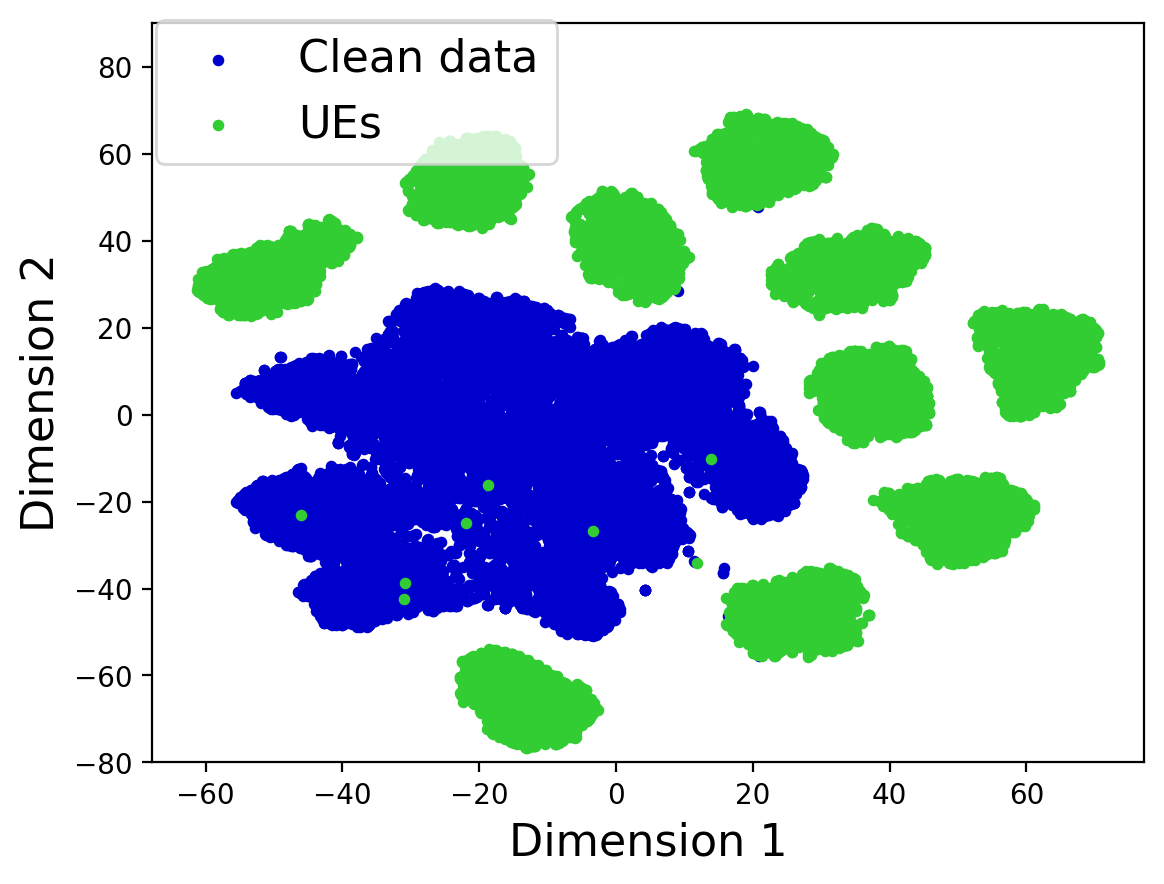}}}
\vspace{-1mm}
\centerline{(b) w/ additional clean data}
\vspace{2mm}
\end{minipage}
\vspace{-3mm}
\caption{
    t-SNE~\cite{tsne} visualizations on CIFAR-10, comparing models trained without and with additional clean data, 
    where the labels for the additional clean data are updated to $y\in[C,2C-1]$.
    Note that UEs are generated by EM, and the shared training data consists of 50\% UEs and 50\% clean data.
    }
    \label{tsne}
\end{figure}

\begin{table}[t]
\caption{Test Accuracy (\%) of models trained on CIFAR-10 with various defensive methods. Dp denote Diffpure~\cite{AVATAR}.}
\vspace{-3mm}
\centering
\scalebox{1.0}{
\setlength{\tabcolsep}{4.5pt}
 \begin{tabular}{ c | c | c  c  c  c  c | c}
\toprule
 Ratio (\%)& Defensive method &EM & OPS& LSP & REM & AR & Mean\\
\midrule
\multirow{3}*{60}& w/o
& 91.86& 91.76 & 91.89& 91.71&91.68&91.78\\
& Dp on all samples & 89.45
&69.55&89.80 & 90.02&89.39&85.64\\
& Dp on detected samples& 92.90
&91.83&92.83 & 91.51&91.79&\textbf{92.17}\\
\midrule
\multirow{3}*{80}& w/o
& 87.94& 87.17 & 88.19& 86.58&87.51&87.47\\
& Dp on all samples & 89.86
&70.39&89.73 & 89.87&89.70&85.91\\
& Dp on detected samples& 91.48
&89.87&90.22 & 91.52&89.97&\textbf{90.61}\\
\bottomrule
\end{tabular}
}
\label{table:detection_for_purification}
\end{table}

\subsection{Detection for Purification}
In this section, we demonstrate the effectiveness of our approach by conducting experiments dedicated to defending against UEs while adhering to detection-defense principles.
It is worth noting that we choose Diffpure~\cite{AVATAR}, which utilizes a diffusion model for purification, as the defense method in these experiments.
As evident in Table~\ref{table:detection_for_purification}, our proposed detection methods prove highly effective in improving defense performance, particularly in scenarios with significant poison ratios. 
{All experiments were conducted on CIFAR-10 dataset.}

\section{Conclusion}
In this paper, we present Iterative Filtering (IF), a robust and efficient detection technique designed to identify a wide range of visually imperceptible UEs.
Our approach leverages a critical property: UEs tend to be acquired by the classifier more rapidly than clean data when training on a partially poisoned dataset.
This difference inspires us to introduce an iterative algorithm for data separation.
Moreover, we highlight that distinguishing between clean and poisoned samples is more effective. 
{Through the introduction of the additional classes and the adoption of iterative refinement,
}
our proposed approach achieves improved effectiveness in classifying these two sample types.
Extensive experiments conclusively demonstrate that our method outperforms state-of-the-art detection approaches when challenged with various types of attacks, datasets, and poison ratios.

\vspace{5mm}
\noindent\textbf{Acknowledgement.} This work was done at Rapid-Rich Object Search (ROSE) Lab, School of Electrical \& Electronic Engineering, Nanyang Technological University. This research is supported in part by the NTU-PKU Joint Research Institute and the DSO National Laboratories, Singapore, under the project agreement No. DSOCL22332.

%
%
%
\bibliographystyle{splncs04}
\bibliography{mybibliography}

\end{document}